\documentclass[conference]{IEEEtran}
\usepackage{graphicx}
\usepackage{booktabs}
\usepackage{multirow}
\usepackage{amsmath,amssymb}
\usepackage{url}
\usepackage{xcolor}
\usepackage{algorithm}
\usepackage{algorithmic}
\usepackage{array}
\usepackage{cite}
\usepackage{graphicx, tikz}
\usetikzlibrary{arrows.meta} 

\usepackage{amsthm}

\newtheorem{remark}{Remark}

\usepackage{tikz}
\usetikzlibrary{arrows.meta,positioning,fit,calc}
\usepackage{subcaption}

\usepackage{float}
\usepackage{needspace}
\title{AI Sessions for Network-Exposed AI-as-a-Service}

\author{
     Mohaned Chraiti \IEEEauthorrefmark{1} and Merve Saimler \IEEEauthorrefmark{2}\\
    \IEEEauthorrefmark{1}Electronics Engineering Department, Sabancı University, Türkiye\\
    \IEEEauthorrefmark{2} Ericsson Research, Türkiye\\
    
    Emails: merve.saimler@ericsson.com, and mohaned.chraiti@sabanciuniv.edu
}

\begin{document}
\maketitle

\begin{abstract}
Cloud-based Artificial Intelligence (AI) inference is increasingly latency- and context-sensitive, yet today’s AI-as-a-Service is typically consumed as an application-chosen endpoint, leaving the network to provide only best-effort transport. This decoupling prevents enforceable tail-latency guarantees, compute-aware admission control, and continuity under mobility. This paper proposes Network-Exposed AI-as-a-Service (NE-AIaaS) built around a new service primitive: the AI Session (AIS)—a contractual object that binds model identity, execution placement, transport Quality-of-Service (QoS), and consent/charging scope into a single lifecycle with explicit failure semantics. We introduce the AI Service Profile (ASP), a compact contract that expresses task modality and measurable service objectives (e.g., time-to-first-response/token, p99 latency, success probability) alongside privacy and mobility constraints. On this basis, we specify protocol-grade procedures for (i) DISCOVER (model/site discovery), (ii) AI PAGING (context-aware selection of execution anchor), (iii) two-phase PREPARE/COMMIT that atomically co-reserves compute and QoS resources, and (iv) make-before-break MIGRATION for session continuity. The design is standard-mappable to Common API Framework (CAPIF) style northbound exposure, ETSI Multi-access Edge Computing (MEC) execution substrates, 5G QoS flows for transport enforcement, and Network Data Analytics Function (NWDAF) style analytics for closed-loop paging/migration triggers.
\end{abstract}


\section{Introduction}
AI inference is increasingly a real-time component of interactive services, spanning conversational assistants, multimodal agents, perception pipelines, and closed-loop control. In these regimes, service quality and safety are shaped by tail behavior rather than averages: time-to-first-response (or first token) and high-quantile end-to-end latency dominate user experience, and mobility adds a separate requirement of continuity. Tail effects are not a corner case; they are a first-order systems phenomenon in large-scale, multi-stage services, where rare but severe delays determine perceived reliability \cite{DeanBarroso2013Tail}. For networked services, this naturally connects to a long line of work on performance guarantees, burstiness, and delay bounding mechanisms \cite{Chang2000SNC,LeBoudecThiran2001NetCalc,WuNegi2003,Neely2010}, as well as statistical tail modeling tools that are explicitly designed to reason about extremes rather than means \cite{Coles2001EVT}.

Despite this, today’s AI-as-a-Service (AIaaS) consumption model remains endpoint-centric: the application selects a model endpoint, often in a remote cloud region, and the network provides best-effort transport. This separation is structural. The network is unaware of the AI-specific objectives that define correctness for the service, while AI infrastructure is unaware of network dynamics that shape tail latency and availability. As a result, there is no enforceable end-to-end contract spanning model choice, compute placement, and transport treatment, and no principled mechanism to preserve service continuity as users move across access and edge domains. This gap is precisely what motivates AIaaS as a network-native capability in emerging AI-native connectivity visions, where service providers are expected to expose and monetize programmable service semantics rather than only raw connectivity \cite{SaimlerSlides2024,ITUIMT2030,EricssonGlobalNetworkAPI2023,GSMAOpenGateway,CAMARAProject}.

This paper argues that the missing primitive is not another placement heuristic, but a session-level service abstraction that makes AI delivery contractible in the same sense that cellular systems made connectivity contractible. 5G already provides the key enforcement levers: differentiated forwarding is enforced at Quality-of-Service (QoS) flow granularity within a (Packet Data Unit (PDU) session through the QoS Flow Identifier (QFI) and associated treatment \cite{TS23501}, while the Policy and Charging Control framework provides an explicit control architecture to admit, steer, and account for such treatment through well-defined functions and interfaces \cite{TS23503}. In parallel, the management trajectory toward intent-driven and closed-loop operation is explicit in standardization efforts on intent-driven management and zero-touch automation \cite{3gppTR28812,etsiZSM002}. The technical question is therefore not whether the substrate exists, but how to turn it into a coherent AI service whose objectives, feasibility, enforcement, and failure semantics are explicit and auditable.

We propose Network-Exposed AI-as-a-Service (NE-AIaaS), in which the service provider provides AI and connectivity as a single service with network-grade semantics. The central abstraction is an AI Session (AIS): a lifecycle object that binds a concrete model identity and version, a concrete execution anchor (edge, regional, or central), a concrete transport treatment (QoS-flow enforcement and steering), and a concrete consent and charging scope into one enforceable contract with diagnosable outcomes. The session is parameterized by an AI Service Profile (ASP), a compact intent descriptor restricted to measurable boundary quantities (for example, time-to-first-response, tail-latency percentiles, completion probability under a timeout), together with feasibility constraints such as privacy/sovereignty scope and mobility class. Under NE-AIaaS, discovery, admission, binding, and migration are control-plane procedures rather than application conventions: the client establishes and maintains a contract whose compliance can be verified and whose violation has deterministic consequences.

The paper makes three contributions. First, it defines the AIS and ASP as minimal contract objects that render AI delivery discoverable, admissible, enforceable, and mobile, while keeping compliance testable at the session boundary. Second, it specifies a standards-mappable architecture that composes exposure, edge execution, analytics, and policy enforcement into explicit lifecycle roles using existing service provider frameworks (Common API Framework, Multi-access Edge Computing, QoS flows, Network Data Analytics Function, and optional radio access nework policy guidance) \cite{3gpp23222capif,etsiMEC003,TS23501}. Third, it derives protocol-grade procedures for discovery, context-aware anchoring, transactional binding of compute and transport, serving with compliance telemetry, and make-before-break migration under mobility, and then presents a standardization and deployment roadmap aligned with ongoing network API ecosystems \cite{GSMAOpenGateway,CAMARAProject}.

\subsection{Paper Outline}
The main objective of this paper is to make AI delivery \emph{contractible}---i.e., jointly admissible over compute and transport resources, enforceable via network primitives, and falsifiable from boundary measurements under mobility. Section~II formalizes the system context and the pass/fail requirements that an enforceable network-exposed AI service must satisfy (Table~\ref{tab:reqs}). Based on these requirements, Section~III develops the minimal contract objects, the AI Service Profile (ASP) and AI Session (AIS), that express service intent, admissibility constraints, and compliance targets in boundary-measurable terms. The missing piece is then the \emph{transactional lifecycle} that can bind compute placement and QoS treatment without partial states; Section~IV completes this by specifying protocol-grade procedures (DISCOVER, AI paging/anchoring, \texttt{PREPARE/COMMIT}, SERVE, and MIGRATION) and mapping them onto standards-mappable enforcement planes. Section~V evaluates the resulting semantics in simulation under load and mobility. Section~VI concludes with implications for standardization and deployment.

\section{System Context and Design Constraints}\label{sec:background}

\subsection{Standard-Mappable Control Primitives}\label{sec:substrate}
NE-AIaaS can take advantage of four service provider-controlled primitives that can be jointly bound into a single service contract: (i) northbound API exposure with explicit onboarding, discovery, and authorization; (ii) edge hosting and lifecycle control for inference runtimes; (iii) enforceable user-plane differentiation at flow granularity; and (iv) analytics interfaces that support feasibility-based admission and closed-loop re-anchoring.

Common API Framework (CAPIF) provides the exposure discipline required for a network-owned API front door, including onboarding, discovery consistency, and security patterns that prevent AIaaS from degenerating into opaque endpoint catalogs \cite{TS23222}. Its direction toward resource-owner authorization (RNAA) is structurally aligned with AIaaS: session binding and data processing may require authorization by the resource owner when privacy-sensitive inputs are processed or premium network treatment is activated \cite{CAPIF_RNAA}. ETSI Multi-access Edge Computing (MEC) provides a reference architecture for placing applications on virtualization infrastructure close to the access network, with management and reference points sufficient to host inference runtimes at the edge and control their lifecycle under load and mobility \cite{ETSI_MEC003}.

On the enforcement plane, 5G system provides differentiated user-plane treatment at the QoS Flow granularity within a PDU Session via QFI and associated QoS characteristics \cite{TS23501}. This is the enforceable handle that makes latency/loss treatment programmable per service flow rather than per best-effort path. The Policy and Charging Control (PCC) framework provides the corresponding admission and authorization logic to activate, deny, and steer such treatment under policy constraints \cite{TS23501}. Finally, NWDAF provides a standardized basis for collecting measurements and exposing analytics to consumers, enabling admission, anchoring, and migration triggers to be derived from measured feasibility rather than static assumptions \cite{TS23288}. Where Radio Access Network (RAN) side constraints must be respected (e.g., steering away from overloaded edge domains), A1 interface principles provide a standards-aligned path for injecting policy guidance and Machine Learning (ML) management information without collapsing the design into proprietary RAN coupling \cite{ETSI_A1GAP}.

\subsection{Gap to an Enforceable Contract: Requirements}\label{sec:gap}
Endpoint-based AIaaS lacks a lifecycle object that binds model identity, compute reservation, and transport enforcement into one auditable contract, and therefore cannot provide tail guarantees under load nor continuity under mobility. Let the end-to-end inference latency for request $r$ decompose as
\begin{equation}\label{eq:latency_decomp}
L_r = L^{\text{RAN}}_r + L^{\text{BH}}_r + L^{\text{Core}}_r + L^{\text{Queue}}_r + L^{\text{Infer}}_r + L^{\text{Return}}_r,
\end{equation}
where $L^{\text{RAN}}_r$ is the access-side (radio) latency (e.g., scheduling and air-interface delay),
$L^{\text{BH}}_r$ is the backhaul latency between access and the execution site,
$L^{\text{Core}}_r$ is the core/transport latency inside the network (routing/forwarding and related processing),
$L^{\text{Queue}}_r$ is the waiting time in the execution queue,
$L^{\text{Infer}}_r$ is the model execution time, and
$L^{\text{Return}}_r$ is the response-path latency back to the requester.
The terms $\{L^{\text{Queue}}_r, L^{\text{Infer}}_r\}$ are execution-side and are dominated by queue saturation and model-dependent runtime, while $\{L^{\text{RAN}}_r, L^{\text{BH}}_r, L^{\text{Core}}_r, L^{\text{Return}}_r\}$ are transport-side and vary with scheduling, congestion, and steering.
A cloud endpoint can elastically scale compute to shape $\{L^{\text{Queue}}_r, L^{\text{Infer}}_r\}$, but it cannot enforce per-flow transport treatment; conversely, the network can enforce per-flow QoS to shape transport-side latency, but it cannot reserve execution capacity or switch model variants when queues saturate. Under this separation, tail constraints such as
\begin{equation}\label{eq:tail_slo}
\Pr[L_r \leq \ell] \geq 1-\epsilon \qquad \text{or} \qquad \text{p99}(L_r) \leq \ell
\end{equation}
are not enforceable unless the system jointly controls both execution-side and transport-side contributions in \eqref{eq:latency_decomp}. Mobility strengthens the requirement: the latency-minimizing execution anchor becomes time-varying across edge domains, and continuity demands re-anchoring with bounded interruption rather than teardown and re-establishment.


The assumed contract and trust model spans four domains: the invoker domain issuing AI requests, the service provider exposure domain performing authentication and authorization/consent checks and session admission/accounting, the execution domain hosting inference runtimes across heterogeneous sites, and the transport domain enforcing QoS and steering. The design constraint is that service intent and authorization scope must be bound to a single lifecycle object (AI Session) so that compute reservations and transport enforcement are consistent, accounting is session-scoped, and revocation has deterministic effect.

We capture the constraints as pass/fail requirements in Table~\ref{tab:reqs}. Failing any requirement implies that the system cannot claim network-exposed AI service semantics since at least one of enforceability, atomicity, measurability, continuity, or diagnosability is missing.

\begin{table*}[t]
\vspace{2pt}
\centering
\caption{NE-AIaaS requirements (pass/fail) and primary enforcing plane (a requirement fails if the capability is absent).}
\label{tab:reqs}
\setlength{\tabcolsep}{4.5pt}
\renewcommand{\arraystretch}{1.05}
\begin{tabular}{p{0.03\linewidth} p{0.65\linewidth} p{0.2\linewidth}}
\hline
Req. & Definition (pass condition) & Primary enforcing plane \\
\hline
R1 & Discoverability: ASP $\rightarrow$ ranked admissible (model, site) candidates with explicit constraints. & Exposure + catalog \\
R2 & Policy-consistent admission: joint feasibility over compute load and transport QoS. & Exposure + policy + analytics \\
R3 & Atomic binding: commit binds compute \emph{and} transport, otherwise rollback (no partial allocation). & Transactional orchestration \\
R4 & Enforceable transport granularity: objectives enforced at QoS-flow (QFI) granularity. & 5G QoS flow/QFI \\
R5 & Compute-aware QoS: contract includes execution-side terms and emits telemetry to measure them. & Execution + telemetry \\
R6 & Mobility continuity: bounded interruption via make-before-break migration. & Orchestration + steering \\
R7 & Consent/authz binding: invocation and data handling remain within resource-owner authorization scope. & Exposure (authz) \\
R8 & Session accounting: usage/charging attributable to an AIS with deterministic scope. & Exposure + charging \\
R9 & Diagnosable failures: explicit cause codes and timers for scarcity/rollback/migration outcomes. & Control plane \\
R10 & Minimal new primitives: composition of exposure/edge/QoS/analytics standards (no proprietary silo). & Architecture \\
\hline
\end{tabular}
\vspace{-0.2cm}
\end{table*}

\begin{remark}
Table~\ref{tab:reqs} implies a small set of lifecycle procedures: discovery to materialize admissible (model, site) candidates (R1), context-aware anchoring to select a feasible execution point under load and mobility risk (R2, R5), transactional co-reservation to enforce atomic binding and diagnosable failures (R3, R9), serving with boundary-measurable telemetry (R4, R5), and make-before-break migration to preserve continuity under mobility (R6). Section~\ref{sec:procedures} specifies these procedures together with the AI Session invariants.
\end{remark}

\section{AI Service Profile and AI Session Semantics}\label{sec:contract}

NE-AIaaS is meaningful only if (i) service intent can be stated in boundary-measurable terms, and (ii) the system exposes a lifecycle object whose validity is unambiguous in time, so that admission, compliance, and continuity can be defined as well-posed operations rather than best-effort behavior. This section formalizes that contract layer through two objects. The {ASP} expresses intent: it specifies measurable service objectives together with admissibility constraints that restrict where and how inference may be executed. The {AI Session (AIS)} is the committed binding that ties an admitted ASP to a concrete serving configuration (model and execution anchor) and to enforceable transport treatment, under explicit authorization and accounting scope.

\subsection{AI Service Profile: What Can Be Measured and What Is Admissible}\label{sec:asp}
An ASP must be falsifiable at the service boundary, which immediately restricts what can be part of the contract. Let $\textsf{TTFB}$ denote time-to-first-response (time-to-first-token under streaming). Let $L$ denote end-to-end inference latency measured at the invoker--service boundary, and let $F_L^{-1}(\alpha)$ denote its $\alpha$-quantile. Reliability is defined through a completion probability together with a hard timeout that fixes the semantics of success. Finally, sustained delivery is captured through an application-defined rate proxy (tokens/s or frames/s). We therefore represent the objective part of the ASP by the parameters
\begin{equation}\label{eq:asp_slo_vector}
\big(\ell_{\textsf{TTFB}},\ \ell_{0.95},\ \ell_{0.99},\ \rho_{\min},\ T_{\max},\ \nu_{\min}\big),
\end{equation}
where $\ell_{\textsf{TTFB}}$ bounds early response, $\ell_{0.95}$ and $\ell_{0.99}$ bound tail latency, $\rho_{\min}$ is a minimum completion probability, $T_{\max}$ is a hard timeout, and $\nu_{\min}$ is a minimum sustained service rate proxy. These are not arbitrary; each term in \eqref{eq:asp_slo_vector} is chosen so that it can be evaluated by boundary telemetry and used as an admission feasibility target.

The ASP must also restrict admissible realizations; otherwise compliance measurements become ill-defined since the system could silently switch to a different serving configuration. Concretely, an ASP declares (a) task modality and interaction mode, so that the admissible model families and I/O pipelines are determined; (b) a resolvable quality tier, so that the model class is not ambiguous; (c) a privacy/sovereignty scope that constrains admissible execution sites, telemetry, and state transfer; (d) a mobility class that determines whether continuity must be provisioned; (e) a cost envelope for admission; and (f) an ordered fallback ladder that is the only admissible degradation path. These constraint terms are treated as protocol information elements, but their role in the paper is semantic: they restrict the feasible set and prevent unobservable changes of the evaluated system.

\subsection{AI Session: Binding the ASP to Enforceable Commitments}\label{sec:ais}
An AIS exists only after an ASP is admitted and bound to concrete resources. The purpose is to eliminate endpoint ambiguity by making three coupled commitments explicit for the session lifetime: which model version serves the session, where inference executes (the anchor site), and how the network treats the session traffic (an enforceable QoS-flow treatment and steering). In addition, the AIS binds the authorization scope implied by the ASP’s privacy constraints and provides a session-scoped accounting handle, so that revocation and charging are lifecycle-consistent rather than endpoint-specific.

Practically, the session maintains a binding record that stores the identifiers needed by the procedures in Section~\ref{sec:procedures}: a session identifier, a reference to the admitted profile (or its digest), the chosen model/version, the chosen anchor site, a routable service endpoint at that site, the QoS-flow enforcement handle (e.g., QFI) and an associated steering handle, a validity lease, an authorization/consent reference consistent with the declared privacy scope, and a charging reference for metering attribution. The specific encoding is not the contribution here; the contribution is that these elements are jointly bound so that admission, rollback, and migration can be defined without partial states or silent reconfiguration.

\subsection{Derived Semantics: Well-Posed Admission, Falsifiable Compliance, and Continuity}\label{sec:invariants}
The methodology follows from a well-posedness requirement: the ASP objectives \eqref{eq:asp_slo_vector} can be interpreted scientifically only if the session admits a time-indexed notion of being ``in contract.'' This requires separating boundary observables (used to evaluate objectives) from control-plane validity (used to define whether the contract is active). Let $v_{\textsf{cmp}}(t)$ indicate that the compute commitment at the chosen anchor is valid at time $t$ (the execution lease exists and is not expired), and let $v_{\textsf{qos}}(t)$ indicate that the enforceable QoS-flow treatment is valid at time $t$. If a session were allowed to persist with only one side valid, then admission would not be a feasibility notion: the same boundary violation could be explained by compute infeasibility (queue collapse) with preferential transport still active, or by transport infeasibility with compute still reserved. The contract would therefore be non-identifiable. The minimal condition that makes admission and rollback well-defined is the coupling
\begin{equation}\label{eq:atomicity}
\mathsf{Committed}(t)\ \Longleftrightarrow\ v_{\textsf{cmp}}(t)\ \wedge\ v_{\textsf{qos}}(t),
\end{equation}
which excludes partial allocation as a representable committed state.

Compliance is meaningful only through boundary-measurable statistics. Over an observation window, let $\widehat{F}_{L}$ denote the empirical distribution of measured end-to-end latency. Tail compliance is evaluated as
\begin{equation}\label{eq:percentile_compliance}
\widehat{F}_{L}^{-1}(0.95)\le \ell_{0.95}, \qquad \widehat{F}_{L}^{-1}(0.99)\le \ell_{0.99},
\end{equation}
and early-response compliance is evaluated by the observed $\textsf{TTFB}$ relative to $\ell_{\textsf{TTFB}}$. Reliability is evaluated through observed completion under the hard timeout $T_{\max}$, which fixes success semantics. This measurability constraint rules out objectives that cannot be falsified at the session boundary.

Authorization and privacy constraints are contract terms, not external policy artifacts, since they restrict admissible sites, telemetry, and state transfer. Let $v_{\sigma}(t)$ indicate validity of the session’s authorization/consent scope at time $t$. Then contract semantics require
\begin{equation}\label{eq:consent}
\neg v_{\sigma}(t)\ \Longrightarrow\ \mathsf{ServeDisabled}(t^+),
\end{equation}
i.e., service cannot remain active past revocation regardless of resource availability. Similarly, the fallback ladder is required to prevent silent changes of the evaluated system: without an admissible degradation set, compliance measurements over time would conflate service quality with hidden switches in model or anchor.

Mobility continuity introduces one further well-posedness constraint: re-anchoring must not create intervals where the contract is undefined. For non-static mobility classes, this is achieved by make-before-break semantics: the new binding becomes committed before the old binding is released, ensuring that the session remains within the domain where \eqref{eq:atomicity} applies throughout migration.

\begin{remark}
Equations \eqref{eq:atomicity}--\eqref{eq:consent} are semantic constraints that make the ASP objectives \eqref{eq:asp_slo_vector} falsifiable and make admission/migration well-posed operations. Section~\ref{sec:procedures} instantiates these semantics through concrete discovery, anchoring, transactional binding, serving, and migration procedures with explicit timers and diagnosable causes.
\end{remark}

\begin{figure*}[t]
  \centering
  \hspace*{-1cm}
  \includegraphics[width=1.1\textwidth]{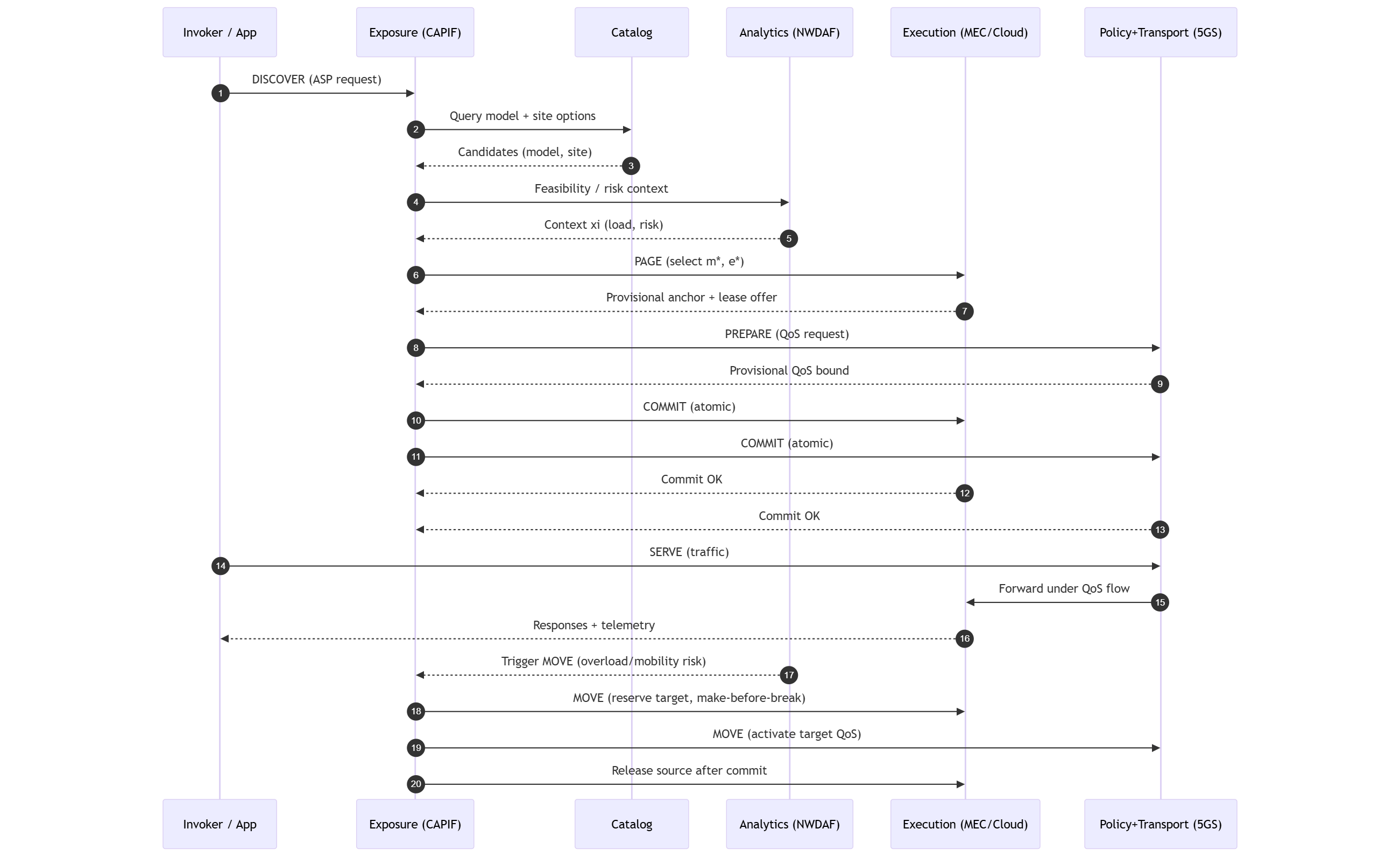}
  \caption{NE-AIaaS end-to-end workflow: discover and anchor a (model,site) binding, atomically co-reserve compute and QoS (\texttt{PREPARE/COMMIT}), serve, and migrate (make-before-break) under risk triggers.}
  \label{fig:architecture}
  \vspace{-0.4cm}
\end{figure*}

\section{NE-AIaaS Architecture and Protocol Procedures}\label{sec:procedures}
This section instantiates the contract semantics of Section~\ref{sec:contract} as an implementable control-plane methodology. The design target is not a descriptive architecture, but a set of cross-plane transactions that make the contract well-posed: feasibility must be materialized explicitly, binding must be atomic across compute and transport, compliance must be falsifiable at the boundary, and continuity must be preserved under mobility. These properties are not optional engineering choices; they are the conditions under which tail guarantees and diagnosable failures can be claimed without ambiguity. Fig.~\ref{fig:architecture} summarizes the end-to-end NE-AIaaS transaction, from discovery and anchoring to atomic commit, serving, and analytics-triggered migration.

\subsection{Architecture as a Composition of Enforceable Planes}\label{sec:arch}
NE-AIaaS composes five logical roles that map to existing standards without introducing a new proprietary network function. The exposure role provides a network-owned northbound entry point with onboarding, discovery discipline, and authorization patterns; CAPIF is the canonical substrate, and its resource-owner authorization direction is aligned with consent-dominant AI sessions when sensitive inputs are processed and premium resources may be triggered \cite{TS23222,CAPIF_RNAA}. A catalog role provides resolvable model identity and admissibility constraints (quality tier, hardware dependency, sovereignty constraints) so that discovery outputs are auditable and do not degenerate into opaque endpoint lists. An execution role provides placement and reservation across heterogeneous sites; ETSI MEC provides the standard-mappable substrate for edge-hosted runtimes and lifecycle management, while regional and central sites expose the same abstract reservation and telemetry hooks \cite{ETSI_MEC003}. A transport role binds session traffic to enforceable QoS-flow treatment and steering; this maps directly to the 5G QoS flow model and to policy and charging control for admitting, denying, and enforcing differentiated treatment \cite{TS23501,TS23501}. An analytics role provides coarse context and risk signals so that admission, anchoring, and migration decisions are driven by measured feasibility rather than static assumptions; NWDAF is the natural standards anchor for analytics exposure and consumption \cite{TS23288}. When anchoring must respect RAN-side constraints without proprietary coupling, A1 policy guidance provides a standards-aligned path for injecting such constraints into the control loop \cite{ETSI_A1GAP}.

The architectural claim is therefore compositional: each plane already exists in modern systems, but NE-AIaaS requires that they be coupled by a small set of transactions derived from the contract semantics.

\subsection{Procedures as Derived Transactions}\label{sec:proc}
The procedure set is dictated by correctness: the NE-AIaaS contract is only meaningful if feasibility is made explicit, binding is atomic across compute and transport, compliance is falsifiable at the session boundary, and continuity is preserved under mobility without teardown. These conditions force five phases: discovery to expose feasible bindings, anchoring to select a serving point under current context, transactional co-reservation to prevent partial allocation from being representable as a committed state, serving to generate boundary evidence for compliance, and make-before-break migration to preserve continuity as feasibility changes over time.

Discovery returns admissible bindings between a concrete model version and an execution anchor after applying the ASP constraints and service provider policy. A candidate binding is represented by $(m,e)$, where $m$ identifies the selected model version and $e$ identifies the execution anchor (edge, regional, or central). To avoid opaque endpoint enumeration, each admissible binding is annotated with predicted boundary quantities that the ASP can constrain: an estimated time-to-first-response $\widehat{T}_{\mathrm{ff}}(m,e)$, an estimated $99$th-percentile end-to-end latency $\widehat{L}_{99}(m,e)$, and an estimated session cost $\widehat{\Gamma}(m,e)$. The discovery output is therefore
\begin{equation}\label{eq:discover_output_revised2}
\mathcal{K}\triangleq \Big\{(m,e):\ \widehat{T}_{\mathrm{ff}}(m,e),\ \widehat{L}_{99}(m,e),\ \widehat{\Gamma}(m,e)\Big\},
\end{equation}
where membership in $\mathcal{K}$ is determined by hard constraints (sovereignty, privacy scope, quality tier, hardware dependencies), while ranking is determined by compliance margin against ASP bounds. Using limits $\ell_{\mathrm{ff}}$ for time-to-first-response and $\ell_{99}$ for tail latency, a natural slack score for a candidate $(m,e)$ is
\begin{equation}\label{eq:discover_margin_revised2}
\Delta(m,e)\triangleq \min\{\ell_{99}-\widehat{L}_{99}(m,e),\ \ell_{\mathrm{ff}}-\widehat{T}_{\mathrm{ff}}(m,e)\}-\lambda\,\widehat{\Gamma}(m,e),
\end{equation}
with policy weight $\lambda\ge 0$. Candidates with $\Delta(m,e)<0$ are predicted to violate at least one bound after accounting for cost policy and are therefore not admissible as compliant choices.

Anchoring cannot be derived from discovery alone since feasibility is context-dependent. Load in the access, backhaul, and execution queues evolves, and mobility changes the relative advantage of anchors over the session horizon. Let $\xi$ denote a coarse service provider-side context summary sufficient to condition feasibility without exposing sensitive details. Anchoring selects a binding $(m^\star,e^\star)\in\mathcal{K}$ by minimizing predicted contract-violation risk expressed directly in the ASP bounds:
\begin{equation}\label{eq:paging_risk_revised2}
\begin{aligned}
&(m^\star,e^\star)=\arg\min_{(m,e)\in\mathcal{K}}
\Big(
w_1\,\widehat{\mathbb{P}}\big[L_{99}>\ell_{99}\,\big|\,m,e,\xi\big]
+
w_2\\ &\,\widehat{\mathbb{P}}\big[T_{\mathrm{ff}}>\ell_{\mathrm{ff}}\,\big|\,m,e,\xi\big]
+
w_3\,\widehat{\mathbb{P}}\big[\text{migration required}\,\big|\,m,e,\xi\big]
\Big),
\end{aligned}
\end{equation}
subject to the hard sovereignty/privacy and cost constraints already enforced in \eqref{eq:discover_output_revised2}. The predictors in \eqref{eq:paging_risk_revised2} are not an algorithmic embellishment; they are the mechanism that ties anchoring to falsifiable outcomes since the events are written in the same boundary quantities that the ASP constrains.

Once $(m^\star,e^\star)$ is selected, the contract forces transactional co-reservation across compute and transport since a committed session must not exist under partial allocation. Let $v_{\mathrm{cmp}}(t)$ denote validity of the compute lease at time $t$ and let $v_{\mathrm{net}}(t)$ denote validity of the enforceable network treatment (QoS-flow binding and steering) at time $t$. Commitment is defined by the equivalence
\begin{equation}\label{eq:commit_semantics_revised2}
\mathsf{Committed}(t)\ \Longleftrightarrow\ v_{\mathrm{cmp}}(t)\wedge v_{\mathrm{net}}(t),
\end{equation}
which forces a two-stage transaction: a provisional stage that obtains both leases and a commit stage that either confirms both or releases both. Without \eqref{eq:commit_semantics_revised2}, tail guarantees are ill-defined since a session could appear established while lacking either compute or enforceable transport treatment.

Determinism requires that each phase execute under an explicit deadline and that failures be attributable to corrective classes rather than collapsed into opaque timeouts. Let $\tau_{\mathrm{disc}},\tau_{\mathrm{page}},\tau_{\mathrm{prep}},\tau_{\mathrm{com}},\tau_{\mathrm{mig}}$ denote the maximum allowed durations for discovery, anchoring, provisional reservation, commit, and migration. The values are policy parameters but the ordering is constrained:
\begin{equation}\label{eq:timer_order_revised2}
\tau_{\mathrm{disc}}\le \tau_{\mathrm{page}}\le \tau_{\mathrm{prep}}\le \tau_{\mathrm{com}},
\qquad
\tau_{\mathrm{mig}}\le \min\{T_{\max},\ \text{lease}\}.
\end{equation}
A compact semantic partition sufficient for diagnosis is
\begin{equation}\label{eq:cause_classes_revised2}
\begin{aligned}
&\mathcal{F}\triangleq \{\text{consent violation},\ \text{policy denial},\ \text{sovereignty violation},\ \\
&\text{model unavailable},\ \text{no feasible binding},\ \text{compute scarcity},\ \\ &\text{QoS scarcity},
\ \text{state transfer failure},\ \text{deadline expiry}\},
\end{aligned}
\end{equation}
where each element implies a distinct remediation path and therefore must not be conflated with others.

After commitment, delivery proceeds over the enforceable QoS-flow treatment and the selected execution anchor, while the runtime produces boundary-measurable telemetry sufficient to evaluate compliance and keep the risk predictors calibrated. Let $T_{\mathrm{ff}}$ denote time-to-first-response and let $Q_L(p)$ denote the $p$-quantile of end-to-end latency. Over a measurement window, the controller observes
\begin{equation}\label{eq:telemetry_revised2}
Z(t)\triangleq \big(\widehat{T}_{\mathrm{ff}}(t),\ \widehat{Q}_L(0.95;t),\ \widehat{Q}_L(0.99;t),\ \widehat{\rho}(t),\ \widehat{q}(t),\ \widehat{\nu}(t)\big),
\end{equation}
where $\widehat{\rho}(t)$ is empirical completion probability under the timeout bound, $\widehat{q}(t)$ is a queue-delay proxy, and $\widehat{\nu}(t)$ is a sustained service-rate proxy. Compliance is evaluated directly against ASP bounds, e.g., $\widehat{T}_{\mathrm{ff}}(t)\le \ell_{\mathrm{ff}}$ and $\widehat{Q}_L(0.99;t)\le \ell_{99}$, which preserves falsifiability by construction.

Mobility and evolving load require make-before-break migration since feasibility changes over time even if the ASP is fixed. A trigger based on predicted violation risk can be written directly against the ASP bounds:
\begin{equation}\label{eq:migration_trigger_revised2}
\widehat{\mathbb{P}}\big[L_{99}>\ell_{99}\,\big|\,\xi\big]\ge \delta
\qquad \text{or} \qquad
\widehat{\mathbb{P}}\big[T_{\mathrm{ff}}>\ell_{\mathrm{ff}}\,\big|\,\xi\big]\ge \delta',
\end{equation}
for policy thresholds $\delta,\delta'$. Migration repeats discovery and anchoring to select a target binding, obtains a provisional co-reservation for the target while the current binding remains committed, then commits the target before releasing the current binding. If state transfer fails or the migration deadline in \eqref{eq:timer_order_revised2} expires, migration aborts while preserving the existing committed service, ensuring that the session never leaves the domain where \eqref{eq:commit_semantics_revised2} holds.


\subsection{Standardization and Deployment Roadmap}\label{sec:roadmap}
The standardization claim is minimal and concrete: the network already contains enforceable primitives for exposure, placement, QoS control, and analytics, but lacks a contract schema and transactional lifecycle semantics that couple those primitives into a falsifiable service. The goal is not to standardize AI implementations, but to standardize the smallest set of artifacts that prevents fragmentation into proprietary AI slices and makes service provider-grade guarantees auditable.

\subsubsection{Standards Mapping and Interoperable Core}\label{sec:fit}
The exposure surface aligns with CAPIF discipline for onboarding, discovery, and authorization, including resource-owner authorization patterns once AI sessions can trigger premium QoS and edge compute \cite{TS23222,CAPIF_RNAA}. Transport enforcement aligns with QoS flows and policy control in 5G, which already provide mechanisms to admit, deny, and steer differentiated treatment \cite{TS23501}. Observability and risk-driven control align with NWDAF-style analytics exposure \cite{TS23288}. Edge placement aligns with MEC hosting and lifecycle management \cite{ETSI_MEC003}. When RAN constraints must influence anchoring without vendor-specific coupling, A1 policy guidance provides a principled interface for injecting constraints into near-real-time control \cite{ETSI_A1GAP}. The contribution of this paper is the coupling logic that makes these planes jointly actionable under a single session contract.

A standards-track outcome should focus on three artifacts. First, an ASP schema with normative semantics that fixes units, aggregation windows, admissibility constraints, and privacy/sovereignty classes, so that discovery and compliance are interoperable across service providers. Second, a small AI-session API surface for establishment, binding, steering, and migration with explicit deadline semantics and cause classes, so that atomicity and diagnosability are preserved across implementations. Third, a compliance profile that specifies how session bindings map onto existing enforcement primitives (QoS-flow treatment, steering, placement) while explicitly avoiding a new monolithic network-function silo.

\subsubsection{Deployment Path and Hard Open Problems}\label{sec:open}
A pragmatic rollout starts in a single-service provider edge domain where exposure, placement, and QoS control fall under one administrative boundary, and then extends to federated edge deployments as discovery and authorization become multi-domain. The hard open problems are truthfulness and safety problems: binding contract terms to measurements so that guarantees are auditable, defining portable state classes so that migration remains safe under privacy constraints, and preserving resource-owner control under monetization pressure. These are the points where a stable semantic core is necessary; predictor implementations and runtime optimizations should remain competitive space.

\section{Simulation Study}\label{sec:sim}

\subsection{Setup}\label{sec:sim_setup}
We evaluate the proposed network-exposed AIaaS (NE-AIaaS) against a conventional endpoint AIaaS baseline using a Monte-Carlo simulator that captures three coupled contributors to inference service delay: (i) server-side queueing due to offered load, (ii) inference execution time variability, and (iii) transport latency variability (best-effort vs.\ QoS-provisioned delivery).  In this section, we use $\rho\in(0,1)$ denotes the normalized {offered-load point} that controls server-side queueing in~(15), i.e., it is a normalized utilization-like parameter used to simulate congestion (higher $\rho$ implies larger $W_q$). 

For each offered-load point $\rho\in(0,1)$, we generate i.i.d.\ end-to-end latency samples.

In the system model, end-to-end inference latency decomposes as in (1),
\[
L_r = L^{\mathrm{RAN}}_r + L^{\mathrm{BH}}_r + L^{\mathrm{Core}}_r + L^{\mathrm{Queue}}_r + L^{\mathrm{Infer}}_r + L^{\mathrm{Return}}_r,
\]
where $\{L^{\mathrm{Queue}}_r, L^{\mathrm{Infer}}_r\}$ are execution-side terms and $\{L^{\mathrm{RAN}}_r, L^{\mathrm{BH}}_r, L^{\mathrm{Core}}_r, L^{\mathrm{Return}}_r\}$ are transport-side terms. In the simulator, we report this decomposition in the aggregated form
\begin{equation}
L = W_q + L_{\mathrm{infer}} + L_{\mathrm{net}},
\end{equation}
where $W_q$ is the compute/server waiting time (corresponding to $L^{\mathrm{Queue}}$), $L_{\mathrm{infer}}$ is the stochastic inference runtime (corresponding to $L^{\mathrm{Infer}}$), and $L_{\mathrm{net}}$ aggregates the transport-side components in (1).

The endpoint baseline represents a fixed cloud endpoint over best-effort transport, where all requests are accepted and can accumulate in the server queue. In contrast, NE-AIaaS models a session-oriented service in which admission is performed via an atomic \texttt{PREPARE/COMMIT} transaction across compute and QoS resources (AI-paging + QoS + atomicity), and only admitted AI sessions are served.

We evaluate two service-level objectives through the ASP consisting of a tail bound $\ell_{99}$ and a hard deadline $T_{\max}$. A served request is declared non-compliant when
\begin{equation}
\text{ASP violation} \Leftrightarrow (L>\ell_{99}) \ \vee\ (L>T_{\max}).
\end{equation}
For endpoint AIaaS, violation probability is computed over all requests (queueing is part of the user-perceived service). For NE-AIaaS, violation probability is computed over admitted sessions only (``served-and-failed''), consistent with session semantics. Finally, we simulate mobility-induced disruption by sweeping the user speed $v$ and computing interruption probability within a fixed session window under two handover mechanisms: (i) teardown/re-establish (baseline), and (ii) make-before-break session continuity (NE-AIaaS).

\subsection{Results}\label{sec:sim_results}
\begin{figure}[t]
  \centering
  \includegraphics[width=0.9\linewidth]{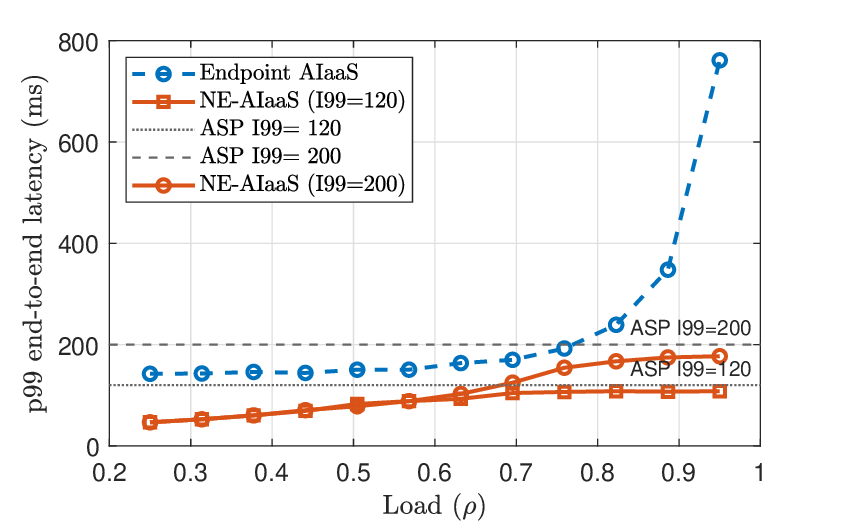}
  \caption{p99 end-to-end latency vs. offered load: NE-AIaaS delays tail collapse via joint compute admission and QoS.}
  \label{fig:p99_load}
    \vspace{-0.1cm}
\end{figure}

Fig.~\ref{fig:p99_load} reports the $99$th-percentile end-to-end latency versus offered load. The endpoint baseline exhibits the expected tail blow-up as $\rho$ approaches saturation, reflecting queue growth and best-effort transport variability. NE-AIaaS maintains substantially lower tail latency over the full operating range because the network participates in selecting an admissible execution/placement option and provisions QoS jointly with compute admission.

\begin{figure}[t]
  \centering
  \includegraphics[width=0.9\linewidth]{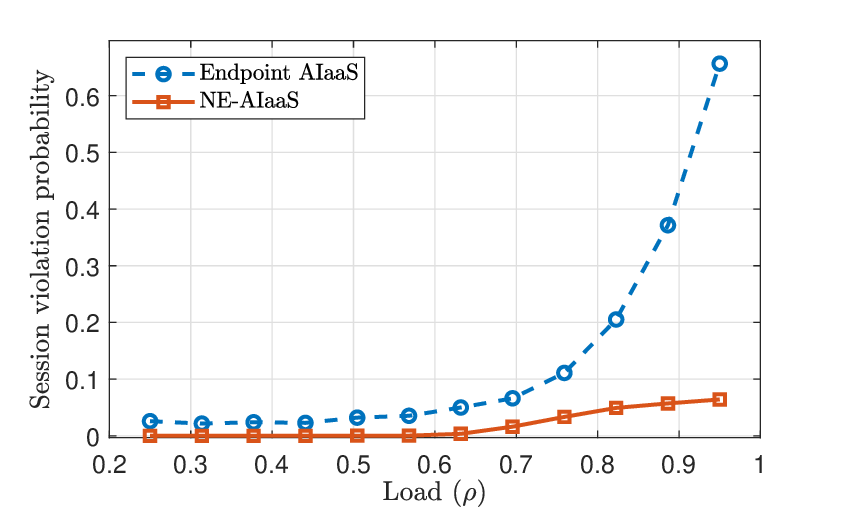}
  \caption{ASP violation probability vs. offered load: NE-AIaaS served-and-failed over admitted sessions.}
  \label{fig:viol_load}
    \vspace{-0.2cm}
\end{figure}

Fig.~\ref{fig:viol_load} shows the ASP violation probability as a function of offered load. For endpoint AIaaS, the violation probability rises sharply near saturation. This stems from the fact that the server queue dominates delay and pushes a growing fraction of requests beyond $\ell_{99}$ and/or $T_{\max}$. In contrast, NE-AIaaS achieves markedly lower violation probability across the load range since admitted sessions benefit from reserved QoS and compute-aware admission.

\begin{figure}[t]
  \centering
  \includegraphics[width=0.9\linewidth]{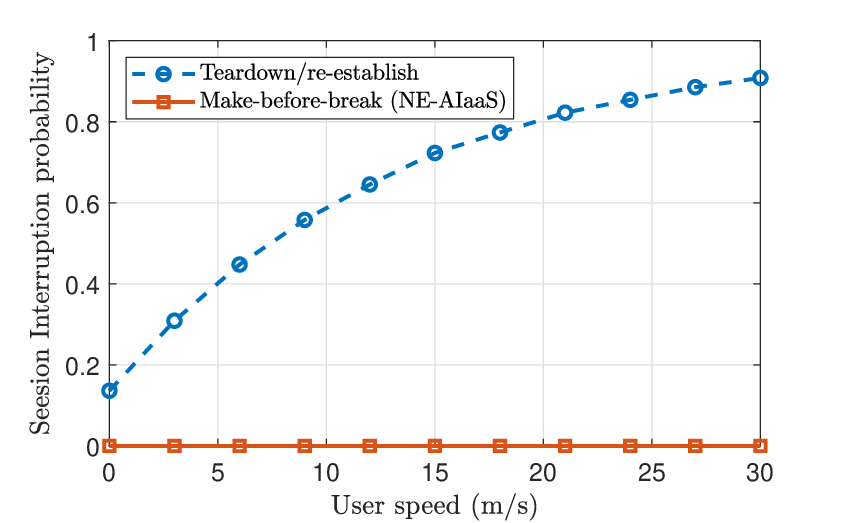}
  \caption{Interruption probability vs. user speed: make-before-break migration preserves continuity versus teardown.}
  \label{fig:mobility}
    \vspace{-0.1cm}
\end{figure}

Fig.~\ref{fig:mobility} evaluates interruption probability versus user speed. Teardown/re-establish induces a rapidly increasing interruption probability with speed due to repeated session teardown and re-setup. In contrast, make-before-break continuity keeps interruption probability close to zero across the entire speed range.

\section{Conclusion}\label{sec:conclusion}
This paper argued that endpoint-centric AIaaS cannot deliver auditable tail-latency guarantees or mobility continuity because compute placement and transport treatment are controlled independently. We proposed Network-Exposed AIaaS (NE-AIaaS), centered on two minimal contract objects: the AI Service Profile (ASP), which restricts intent to boundary-measurable objectives and admissibility constraints, and the AI Session (AIS), which binds a concrete model/version, execution anchor, enforceable QoS treatment, and consent/charging scope into a single lifecycle with explicit failure causes. On this basis, we specified protocol-grade procedures for discovery, context-aware anchoring (AI paging), atomic \texttt{PREPARE/COMMIT} co-reservation of compute and QoS, compliance telemetry, and make-before-break migration. A simulation study illustrated that contract-aware admission and joint network--compute control delay tail collapse under load and reduce served-and-failed ASP violations while preserving continuity under mobility. We conclude that standardizing the ASP/AIS semantic core and transactional lifecycle is a practical path to interoperable, provider-grade AI services that remain measurable, diagnosable, and enforceable.

\bibliographystyle{IEEEtran}
\bibliography{IEEEabrv, refs}

@STRING{IEEE_J_WCOM       = "{IEEE} Trans. Wireless Commun."}

@techreport{TS23222,
  author      = {{ETSI} and {3GPP}},
  title       = {{ Common API Framework for 3GPP Northbound APIs (3GPP TS 23.222 Release 18)}},
  institution = {ETSI},
  year        = {2025},
  month       = jan,
  url         = {https://www.etsi.org/deliver/etsi_ts/123200_123299/123222/18.07.00_60/ts_123222v180700p.pdf},
  note        = {Accessed: 2026-01-28}
}

@misc{CAPIF_RNAA,
  author       = {{3GPP}},
  title        = {{CAPIF Framework -- RNAA (Resource Owner NAA) Technology Page}},
  howpublished = {3GPP Technologies},
  year         = {2023},
  month        = jul,
  url          = {https://www.3gpp.org/technologies/rnaa},
}

@techreport{TS23501,
  author      = {{ETSI} and {3GPP}},
  title       = {{System architecture for the 5G System (3GPP TS 23.501 Release 18)}},
  institution = {ETSI},
  year        = {2025},
  month       = jul,
  url         = {https://www.etsi.org/deliver/etsi_ts/123500_123599/123501/18.10.00_60/ts_123501v181000p.pdf},
}

@techreport{TS23503,
  author      = {{ETSI} and {3GPP}},
  title       = {Policy and charging control framework for the 5G System (5GS); Stage 2 (3GPP TS 23.503 Release 18)},
  institution = {European Telecommunications Standards Institute (ETSI)},
  year        = {2025},
  month       = apr,
  url         = {https://www.etsi.org/deliver/etsi_ts/123500_123599/123503/18.09.00_60/ts_123503v180900p.pdf},
}

@techreport{TS23288,
  author      = {{ETSI} and {3GPP}},
  title       = { Architecture enhancements for {5G System (5GS)} to support network data analytics services ({3GPP TS 23.288 Release 18})},
  institution = {European Telecommunications Standards Institute (ETSI)},
  year        = {2025},
  month       = jul,
  url         = {https://www.etsi.org/deliver/etsi_ts/123200_123299/123288/18.10.00_60/ts_123288v181000p.pdf},
}

@techreport{ETSI_MEC003,
  author      = {{ETSI}},
  title       = {{Multi-access Edge Computing (MEC); Framework and Reference Architecture}},
  institution = {European Telecommunications Standards Institute (ETSI)},
  year        = {2025},
  month       = may,
  url         = {https://www.etsi.org/deliver/etsi_gs/mec/001_099/003/04.01.01_60/gs_mec003v040101p.pdf},
}

@techreport{ETSI_A1GAP,
  author      = {{ETSI}},
  title       = {{A1 interface: General Aspects and Principles (O-RAN.WG2.A1GAP-R004-v04.00)}},
  institution = {European Telecommunications Standards Institute (ETSI)},
  year        = {2025},
  month       = may,
  url         = {https://www.etsi.org/deliver/etsi_ts/103900_103999/103983/04.00.00_60/ts_103983v040000p.pdf},
}

@misc{SaimlerSlides2024,
  author       = {Saimler, Merve},
  title        = {The Dawn of {AI-Native} Networks and Transforming Connectivity with {AI-as-a-Service}},
  year         = {2024},
  month        = dec,
  url          = {https://share.google/4eWWPqk0XjKN3vFzz}
}

@book{Neely2010,
  author    = {Neely, M. J.},
  title     = {Stochastic Network Optimization with Application to Communication and Queueing Systems},
  publisher = {Morgan \& Claypool},
  year      = {2010}
}

@article{WuNegi2003,
  author  = {Wu, D. and Negi, R.},
  title   = {Effective Capacity: A Wireless Link Model for Support of Quality of Service},
  journal =IEEE_J_WCOM, 
  year    = {2003},
  volume  = {2},
  number  = {4},
  pages   = {630--643}
}

@misc{EricssonGlobalNetworkAPI2023,
  author       = {Ericsson},
  title        = {Global network {API} platform to monetize {5G}},
  year         = {2023},
url={https://www.ericsson.com/en/reports-and-papers/white-papers/global-network-api-platform-to-monetize-5g},
}

@techreport{3gpp23222capif,
  author       = {{3GPP}},
  title        = {{Common API Framework for 3GPP Northbound APIs (CAPIF)}},
  institution  = {ETSI / 3GPP},
  year         = {2024},
  url          = {https://www.etsi.org/deliver/etsi_ts/123200_123299/123222/15.07.00_60/ts_123222v150700p.pdf}
}

@techreport{etsiMEC003,
  author       = {{ETSI}},
  title        = {{Multi-access Edge Computing (MEC); Framework and Reference Architecture}},
  institution  = {ETSI},
  number       = {ETSI GS MEC 003},
  year         = {2022},
  note         = {v3.2.1},
  url          = {https://www.etsi.org/deliver/etsi_gs/MEC/001_099/003/03.02.01_60/gs_mec003v030201p.pdf}
}

@techreport{etsiZSM002,
  author       = {{ETSI}},
  title        = {{Zero-touch network and Service Management (ZSM); Reference Architecture}},
  institution  = {ETSI},
  number       = {ETSI GS ZSM 002},
  year         = {2019},
  url          = {https://www.etsi.org/deliver/etsi_gs/ZSM/001_099/002/01.01.01_60/gs_ZSM002v010101p.pdf}
}

@misc{GSMAOpenGateway,
  author       = {{GSMA}},
  title        = {{GSMA Open Gateway: State of the Market, H1 2024}},
  year         = {2024},
  note         = {GSMA Intelligence report},
  url          = {https://www.gsmaintelligence.com/research/research-file-download?file=200624-Open-Gateway-State-of-the-Market.pdf&id=79791577}
}

@misc{CAMARAProject,
  author       = {{Linux Foundation CAMARA Project}},
  title        = {{CAMARA: Open Source Project for Telco Network APIs}},
  year         = {2026},
  note         = {Project repositories and governance},
  url          = {https://github.com/camaraproject}
}

@misc{ITUIMT2030,
  author       = {{ITU-R}},
  title        = {{IMT towards 2030 and beyond (IMT-2030)}},
  year         = {2023},
  note         = {ITU-R WP 5D portal page},
  url          = {https://www.itu.int/en/ITU-R/study-groups/rsg5/rwp5d/imt-2030/pages/default.aspx}
}

@techreport{3gppTR28812,
  author       = {{3GPP}},
  title        = {{Study on scenarios for Intent driven management services for mobile networks}},
  institution  = {3GPP},
  number       = {3GPP TR 28.812},
  year         = {2018},
  url          = {https://www.3gpp.org/DynaReport/28812.htm}
}

@article{DeanBarroso2013Tail,
  author  = {Jeffrey Dean and Luiz Andr{\'e} Barroso},
  title   = {The Tail at Scale},
  journal = {Communications of the ACM},
  volume  = {56},
  number  = {2},
  pages   = {74--80},
  year    = {2013},
  doi     = {10.1145/2408776.2408794}
}

@book{Chang2000SNC,
  author    = {Cheng-Shang Chang},
  title     = {Performance Guarantees in Communication Networks},
  publisher = {Springer},
  year      = {2000}
}

@book{LeBoudecThiran2001NetCalc,
  author    = {Jean-Yves Le Boudec and Patrick Thiran},
  title     = {Network Calculus: A Theory of Deterministic Queuing Systems for the Internet},
  publisher = {Springer},
  year      = {2001}
}

@book{Coles2001EVT,
  author    = {Stuart Coles},
  title     = {An Introduction to Statistical Modeling of Extreme Values},
  publisher = {Springer},
  year      = {2001}
}
\end{document}